

\message{   *** Please respond `b' to the query! ***     }

\input harvmac.tex
%
\global\newcount\mthno \global\mthno=1
\global\newcount\mexno \global\mexno=1
\global\newcount\mquno \global\mquno=1
\def\newsec#1{\global\advance\secno by1\message{(\the\secno. #1)}
\global\subsecno=0\xdef\secsym{\the\secno.}\global\meqno=1\global\mthno=1
\global\mexno=1\global\mquno=1
\bigbreak\medskip\noindent{\bf\the\secno. #1}\writetoca{{\secsym} {#1}}
\par\nobreak\medskip\nobreak}
\xdef\secsym{}
\global\newcount\subsecno \global\subsecno=0
\def\subsec#1{\global\advance\subsecno by1\message{(\secsym\the\subsecno. #1)}
\bigbreak\noindent{\it\secsym\the\subsecno. #1}\writetoca{\string\quad
{\secsym\the\subsecno.} {#1}}\par\nobreak\medskip\nobreak}
\def\appendix#1#2{\global\meqno=1\global\mthno=1\global\mexno=1%
\global\mquno=1
\global\subsecno=0
\xdef\secsym{\hbox{#1.}}
\bigbreak\bigskip\noindent{\bf Appendix #1. #2}\message{(#1. #2)}
\writetoca{Appendix {#1.} {#2}}\par\nobreak\medskip\nobreak}
%
%
\def\thm#1{\xdef #1{\secsym\the\mthno}\writedef{#1\leftbracket#1}%
\global\advance\mthno by1\wrlabeL#1}
\def\que#1{\xdef #1{\secsym\the\mquno}\writedef{#1\leftbracket#1}%
\global\advance\mquno by1\wrlabeL#1}
\def\exm#1{\xdef #1{\secsym\the\mexno}\writedef{#1\leftbracket#1}%
\global\advance\mexno by1\wrlabeL#1}
%
%
\def\ref{\the\refno\nref}
\def\nref#1{\xdef#1{\the\refno}\writedef{#1\leftbracket#1}%
\ifnum\refno=1\immediate\openout\rfile=refs.tmp\fi
\global\advance\refno by1\chardef\wfile=\rfile\immediate
\write\rfile{\noexpand\item{[#1]\ }\reflabeL{#1\hskip.31in}\pctsign}\findarg}
\def\bref{\nref}
%

%
%
%

\def\p{\partial}
%
%
\def\al{\alpha} \def\be{\beta} \def\ga{\gamma}  
\def\de{\delta}  \def\De{\Delta} 
  \def\et{\eta} 
    
\def\la{\lambda} \def\La{\Lambda} \def\rh{\rho} 
    
\def\ph{\phi}     
\def\ps{\psi}     
%
%

%
 
%
%
  \def\cF{{\cal F}} 
   
  \def\cW{{\cal W}} 
%
%
%
\def\lefthook{{\vrule height5pt width0.4pt depth0pt}}
\def\righthook{{\vrule height5pt width0.4pt depth0pt}}
\def\leftrighthookfill{$\mathsurround=0pt \mathord\lefthook
     \hrulefill\mathord\righthook$}
\def\underhook#1{\vtop{\ialign{##\crcr$\hfil\displaystyle{#1}\hfil$\crcr
      \noalign{\kern-1pt\nointerlineskip\vskip2pt}
      \leftrighthookfill\crcr}}}
%
%


\def\dCo{{\rm h}^{\vee}}

\def\ie{{\it i.e.\ }}
\def\eg{{\it e.g.\ }}

\def\ZZ{Z\!\!\!Z}               
\def\NN{I\!\!N}                 %
\def\RR{I\!\!R}                 
\def\CC{I\!\!\!\!C}             %
\def\bfg{{\bf g}}
\def\bfh{{\bf h}}
\def\bfnm{{\bf n}_-}
\def\bfnp{{\bf n}_+}



\def\mapright#1{\smash{\mathop{\longrightarrow}\limits^{#1}}}


\hfuzz=5pt

\def\cS{{\cal S}}
\def\Lap{{\La^{(+)}}} \def\Lam{{\La^{(-)}}}
\def\wW{{\widehat{W}}}
\def\det{{\rm det\,}}
\def\bgh#1{b^{[#1]}} \def\cgh#1{c^{[#1]}}
\def\gh#1{{\rm gh}(#1)}
\def\whg{{\widehat{\bfg}}}
\def\cF{F}

\def\hep#1{{\tt hepth@xxx/{#1}}}

\def\PL{Phys.\ Lett.\ }
\def\NP{Nucl.\ Phys.\ }
\def\CMP{Comm.\ Math.\ Phys.\ }
\def\MPL{Mod.\ Phys.\ Lett.\ }
\def\IJMP{Int.\ J.\ Mod.\ Phys.\ }

\bref\TM{
J.~Thierry-Mieg, \PL {\bf 197B} (1987) 368.}

\bref\DDR{
S.R.~Das, A.~Dhar and S.K.~Rama, \MPL {\bf A6} (1991) 3055;
\IJMP {\bf A7} (1992) 2295; S.K.~Rama, \MPL {\bf A6} (1991) 3531.}

\bref\Popetala{
C.N.~Pope, L.J.~Romans and K.S.~Stelle, \PL {\bf 268B} (1991) 167;
\PL {\bf 269B} (1991) 287;
C.N.~Pope, L.J.~Romans, E.~Sezgin and K.S.~Stelle, \PL {\bf 274B} (1992)
298;
H.~Lu, C.N.~Pope, S.~Schrans and K.W.~Xu, \NP {\bf B385} (1992) 99;
H.~Lu, C.N.~Pope, S.~Schrans and X.J.~Wang, \NP {\bf B379} (1992) 47.}

\bref\BLNWa{
M.~Bershadsky, W.~Lerche, D.~Nemeschansky and N.P.~Warner,
\PL {\bf 292B} (1992) 35.}

\bref\BSSBG{
E.\ Bergshoeff, A.\ Sevrin and X.\ Shen, {\it A derivation of
the BRST operator for noncritical W strings}, preprint
UG-8/92, \hep{9209037}; J.\ de Boer and  J.\ Goeree, {\it KPZ analysis for
$W_3$ gravity}, preprint THU-92/34, \hep{9211108}.}

\bref\Popetalb{
C.N.~Pope, E.~Sezgin, K.S.~Stelle and X.J.~Wang, {\it Discrete states
in the $W_3$ string}, CTP~TAMU-64/92, \hep{9209111};
P.~West, {\it On the spectrum, no ghost theorem and modular invariance
of $W_3$ strings}, KCL-TH-92-7, \hep{9212016};
H.~Lu, B.E.W.~Nilsson, C.N.~Pope, K.S.~Stelle and P.C.~West,
{\it The low-level spectrum of the $W_3$ string}, CTP TAMU-70/92,
\hep{9212017};
H.~Lu, C.N.~Pope, S.~Schrans and X.J.~Wang, {\it On the spectrum and
scattering of $W_3$ strings}, CTP~TAMU-4/93, \hep{9301099};
{\it The interacting $W_3$ string}, CTP~TAMU-86/92, \hfil\break
\hep{9212117}.}

\bref\BLNWb{
M.~Bershadsky, W.~Lerche, D.~Nemeschansky and N.P.~Warner,
{\it Extended $N=2$ superconformal structure of gravity and
$W$-gravity coupled to matter}, CALT-68-1832, \hep{9211040}.}

\bref\Wi{
E.~Witten, \NP {\bf B373} (1992) 187;
E.~Witten and B.~Zwiebach, \NP {\bf B377} (1992) 55.}

\bref\BMPd{
P.~Bouwknegt, J.~McCarthy and K.~Pilch, in preparation.}

\bref\Se{
N.~Seiberg, Prog. Theor. Phys. Suppl. {\bf 102} (1990) 319.}

\bref\BS{
P.~Bouwknegt and K.~Schoutens, Phys. Rep. {\bf 223} (1993) 183.}

\bref\TK{
A.~Tsuchiya and Y.~Kanie, Publ. RIMS, Kyoto Univ. {\bf 22} (1986)
259.}

\bref\Fr{
E.~Frenkel, \PL {\bf 285B} (1992) 71.}

\bref\BMPb{
P.~Bouwknegt, J.~McCarthy and K.~Pilch, {\it Some aspects of free field
resolutions in $2D$ CFT with application to the quantum Drinfeld-Sokolov
reduction}, in proc. ``Strings and Symmetries 1991,'' eds. N.~Berkovitz
et.\ al., World Scientific, 1991.}

\bref\FKW{
E.~Frenkel, V.G.~Kac and M.~Wakimoto, \CMP {\bf 147} (1992) 295.}

\bref\Wa{
G.M.T.~Watts, \NP {\bf B326} (1989) 648; erratum, \NP {\bf B336}
(1990) 720.}

\bref\BMPc{
P.~Bouwknegt, J.~McCarthy and K.~Pilch, {\it Semi-infinite cohomology
in conformal field theory and 2d gravity}, CERN-TH.6646/92,
\hep{9209034}.}

\bref\Knapp{
A.\ Knapp, {\it Lie groups, Lie algebras, and cohomology,} Princeton
University Press, Princeton (1988).}

\bref\Hu{
J.E.~Humphreys, {\it Reflection groups and Coxeter groups}, Cambridge
University Press, Cambridge, 1990.}

\bref\BMPa{
P.~Bouwknegt, J.~McCarthy and K.~Pilch, \CMP {\bf 145} (1992) 541.}

\bref\LZs{
B.H.~Lian and G.J.~Zuckerman, \PL {\bf 254B} (1991) 417;
\PL {\bf 266B} (1991) 21; \CMP {\bf 145} (1992) 561.}

\bref\MMMO{
A.~Marshakov, A.~Mironov, A.~Morozov and M.~Olshanetsky, {\it $c=r_G$
theories of $W_G$-gravity: the set of observables as a model of simply
laced $G$}, FIAN/TD-02/92, \hep{9203044}.}

\bref\Zh{
D.B.~Zhelobenko, {\it Compact Lie groups and their representations},
Providence, RI: American Mathematical Society, 1973.}

\bref\It{
K.~Itoh, {\it $SL(2,\RR)$ current algebra and spectrum in
two dimensional gravity}, CTP-TAMU-42/91.}

\bref\Sa{
V.~Sadov, {\it On the spectra of $\widehat{sl}(N)_k/
\widehat{sl}(N)_k$-cosets and $W_N$ gravities I}, HUTP-92/A055,
\hep{9302060}.}

%
%

\line{}
\vskip2cm

\centerline{\titlefont SEMI-INFINITE COHOMOLOGY OF $\cW$-ALGEBRAS}
\vskip2cm

\centerline{Peter Bouwknegt$^1$
\footnote{$^\dagger$}{Supported by the Packard Foundation.},
Jim McCarthy$^2$ \footnote{$^\ddagger$}{Supported
by the Australian Research Council.}
and Krzysztof Pilch$^1$
\footnote{$^*$}{Supported in part by the
Department of Energy Contract \#DE-FG03-84ER-40168.}} \bigskip

\centerline{\sl $^1$ Department of Physics and Astronomy }
\centerline{\sl  University of Southern California}
\centerline{\sl Los Angeles, CA~90089-0484, USA}
\medskip

\centerline{\sl $^2$ Department of Physics and Mathematical Physics}
\centerline{\sl University of Adelaide, Adelaide, SA~5001, Australia}
\vskip1.5cm

\centerline{\bf Abstract}\smallskip

\noindent We generalize some of the standard
homological techniques to $\cW$-algebras, and
compute the semi-infinite cohomology
of the $\cW_3$ algebra on a variety of modules. These
computations provide physical states in $\cW_3$ gravity coupled to
$\cW_3$ minimal models and to two free scalar fields.


\vfil
\line{USC-93/11\hfil}
\line{ADP-93-200/M15\hfil}
\line{{\tt hepth@xxx/9302086}\hfil February 1993}

\eject


\newsec{Introduction}

An outstanding problem in $\cW$-gravity
 is the computation of physical states. In particular, this has been
studied as an application of BRST
cohomology [\TM-\BLNWb].
One of the ultimate goals is to achieve a better understanding
of (quantum) $\cW$-geometry through the computation of the ground ring and
its associated symmetries [\Wi].  From a mathematical point of view,
the semi-infinite cohomology of a
$\cW$-algebra is interesting since some standard techniques for
computing Lie algebra cohomology no longer apply due to nonlinearity
of the algebra.

In this letter we will summarize various results on the computation
of the BRST cohomology of $\cW$-gravity coupled to $\cW$-matter.
Details and additional results will appear elsewhere [\BMPd].
Throughout the paper we will have the $\cW_3$ algebra in mind, although
some of the results will be formulated for more general $\cW$-algebras
while others can be generalized straightforwardly. We have chosen an
approach that remains entirely within the context of $\cW$-algebras. It
is not inconceivable, though, that our results can also be deduced
from analogous results for affine Lie algebras by quantum Drinfeld-Sokolov
reduction.

In Section 2, apart from introducing notation,
we establish some general results on the structure
of $\cW$-algebra modules. These results
are needed for the cohomology computation,
but are also interesting in their own right.
In Section 3 we will calculate the cohomology of the
BRST operator recently constructed in [\BLNWa], on the
product of an irreducible $\cW_3$ minimal model module and a two-scalar
Fock module (\ie `$\cW_3$ gravity coupled to a $\cW_3$ minimal model')
for the states which satisfy the analogue of the Seiberg bound
in ordinary gravity [\Se].
By specializing these results to the identity
module ($c^M=0$), where the BRST-operator reduces to the one constructed
in [\TM], we find the physical states for `pure $\cW_3$ gravity.'
In Section 4 we present some results concerning the cohomology of
the BRST-operator on the product of two
two-scalar Fock spaces (\ie `$\cW_3$ gravity coupled
to two free scalar fields' or the `$d=(2,2)$ $\cW_3$ string').

\newsec{General results}

The results in this section apply to
the $\cW$-algebras $\cW[\bfg]$ based on a simple simply-laced Lie
algebra $\bfg$ of rank $\ell$.
We will use the notations $W$ for the Weyl group
of $\bfg$, $\wW$ for the Weyl group of the (untwisted) affine Lie
algebra $\whg$, $P_+$ for the set of dominant integral weights of
$\bfg$, $\De_+$ for the set of positive roots of $\bfg$,
$Q_+=\ZZ_+\De_+$ for the positive root lattice of $\bfg$ and
$\ell(w)$ for the length of a Weyl group element $w\in W$ (or $w\in\wW$).
For background material on $\cW$-algebras and a
complete list of notations we refer the reader to the recent review [\BS]
and references therein.

The algebra $\cW[\bfg]$ is generated by $L_n^{(i)}$, where $n\in\ZZ$
and $i$ runs over the orders of the  independent Casimirs of $\bfg$.
In particular, $L_n=L_n^{(2)}$ are the generators of the Virasoro
subalgebra.
The Verma module $M(h^{(i)},c)$
is defined, as usual, as the module induced from
a highest weight vector $|h^{(i)}\rangle$ by the
$ L_n^{(i)}$, $n<0$,
and is labelled by the $L_0^{(i)}$ eigenvalues
$h^{(i)}$ on the highest weight vector.

Let $\cF(\La,\al_0)$ denote the Fock space of $\ell$ scalar fields
$\ph^k(z)$, normalized such that $\ph^k(z)\ph^l(w) =
-\de^{kl}\ln(z-w)$,
and coupled to a background charge $\al_0\rh$, where $\rh$ is
the principal vector of the Lie algebra $\bfg$.
The Fock space vacuum $|\La\rangle$
is labelled by a vector $\La$ in the weight lattice of
$\bfg$ such that $p^k|\La\rangle =\La^k |\La\rangle$.

A realization of $\cW[\bfg]$ on the Fock space
$\cF(\La,\al_0)$ of $\ell$ scalar fields can be constructed by
means of the quantum Drinfeld-Sokolov reduction. In particular, we
have the following expression for the stress-energy tensor
\eqn\eqBa{
T(z)  = -\half (\p\ph(z)\cdot \p\ph(z)) - i\al_0 \rh\cdot\p^2
\ph(z)\,.
}
It generates a Virasoro subalgebra of central charge $c=\ell -12\al_0^2
|\rh|^2$, while
\eqn\eqBb{
h(\La) \equiv h^{(2)} (\La) = \half (\La,\La+2\al_0\rh)\,.
}
The free field realization also induces homomorphisms
\eqn\eqBz{
\matrix{
M(\La,c)& \mapright{\imath'}& F(\La,\al_0) &\mapright{\imath''}
&M(\La,c)^*\cr}\,,
}
where we have written $M(\La,c)$ instead of $M(h^{(i)}(\La),c)$, and
$M(\La,c)^*$ is the contragradient Verma module.
It should be kept in mind, though, that the map $\La\to (h^{(i)}(\La))$
is not 1--1 but rather $(h^{(i)}(\La)) = (h^{(i)}(\La'))$ iff
$\La+\al_0\rh = w(\La'+\al_0\rh)$ for some Weyl group element $w\in W$.

Let $M_N(\La,c)$ and $\cF_N(\La,\al_0)$ denote the subspaces at
energy level $L_0 = h(\La) + N$, and let $\{w_i\}$ and $\{v_i\}$ denote
$p_\ell(N)$-dimensional bases of $M_N(\La,c)$
and $\cF_N(\La,\al_0)$, respectively.
The maps $\imath'$ and $\imath''$ can be analyzed through
the  determinants $\cS'_N(\La,\al_0)$ and $\cS''_N(\La,\al_0)$
of the corresponding Shapovalov forms,
\eqn\eqBc{
\cS'_N(\La,\al_0) =\det{( \vev{v_i|\imath'(w_j)})}\,,\qquad
\cS''_N(\La,\al_0) =  \cS'_N( w_0(\La+\al_0\rh)-\al_0\rh, \al_0)\,,
}
where $\vev{\,\cdot\, |\, \cdot\, }$
denotes the (canonical) form on $\cF(\La,\al_0)$. We have used the
isomorphism $F(\La,\al_0)^* \cong F(w_0(\La+\al_0\rh)-\al_0\rh,
\al_0)$, where $w_0$ is the longest element in $W$.
Defining $\al_\pm$ by $\al_0 = \al_+ + \al_-,\,\al_+\al_-=-1$ we have
\thm\thBa
\proclaim Theorem \thBa.
$$ \eqalign{
\det \cS'_N(\La,\al_0) & \sim \prod_{\al\in\De_+}\
\prod_{\scriptstyle r,s\in\NN \atop \scriptstyle rs\leq N}
\left( (\La+\al_0\rh,\al) - (r\al_+ + s\al_-) \right)^{p_\ell(N-rs)}\,, \cr
\det \cS''_N(\La,\al_0) & \sim \prod_{\al\in\De_+}\
\prod_{\scriptstyle r,s\in\NN \atop \scriptstyle rs\leq N}
\left( (\La+\al_0\rh,\al) + (r\al_+ + s\al_-)\right)^{p_\ell(N-rs)}\,. \cr}
$$
\par

The proof parallels the one given in [\TK] for the Virasoro algebra.
The product of the two determinants is, of course, proportional to
the usual Kac determinant of the Shapovalov form on $M(\La)$
(see \eg [\BS] and references therein).

As an immediate consequence of Theorem \thBa\ we have
\thm\thBb
\proclaim Corollary \thBb.
\item{(a)}
$
\cF(\La,\al_0) \ \cong\ \cases{
M(\La,c) & if  $(\La+\al_0\rh,\al)\notin
(\NN\al_+ + \NN\al_-)$ for all $\al\in\De_+\,,$\cr
M(\La,c)^* & if $(\La+\al_0\rh,\al)\notin
-(\NN\al_+ + \NN\al_-)$ for all $\al\in\De_+\,.$ \cr}
$\hfil\break
\noindent
In particular, if $(\La+\al_0\rh,\al) \notin (\NN\al_+ + \NN\al_-)$
for all $\al\in\De$, then $M(\La,c)$ (and thus also $\cF(\La,\al_0)$)
is irreducible.
\item{(b)}
For $\al_0{}^2 \leq-4$ or, equivalently,
$c\geq c_{crit}-\ell =  \ell + 48 |\rh|^2$ we have
$$
\cF(\La,\al_0) \cong \cases{
M(\La,c) & for $i(\La+\al_0\rh) \in \et D_+\,,$ \cr
M(\La,c)^* & for $-i(\La+\al_0\rh) \in \et D_+\,,$ \cr}
$$
where $D_+ = \{ \la\in \bfh_{\,\CC}^*\ | \ (\la,\al)\in\RR_+,\ \forall
\al\in\De_+\}$ denotes the fundamental Weyl chamber, and $\et = {\rm sign}
(-i\al_0)$.

Part (a) of the theorem follows from the absence of zeros in the corresponding
determinants, whilst part (b) follows from (a) by observing that
$-i\al_\pm\in\et\RR_+$
for $c\geq c_{crit} -\ell$ (see [\Fr] for $\bfg = sl(2)$).

The embedding structure of the Fock spaces $\cF(\La,\al_0)$ follows,
in principle, from Theorem \thBa. It is
quite complicated in general, except for $c=\ell$  (\ie $\al_0=0$)
where we have the following result
\thm\thBc
\proclaim Theorem \thBc. If $w\in W$ such that $w\La\in P_+$ then
\eqn\eqBd{
\cF (\La) = \bigoplus_{\scriptstyle \be\in Q_+\atop \scriptstyle
w\La+\be\in P_+}
m(w\La;\be)\,L(w\La+\be)\,,
}
where, for  $\La\in P_+$ and $\be\in Q_+$, the multiplicity $m(\La;\be)$
(with which the $c=\ell$ irreducible $\cW$-module $L(\La+\be)$
occurs in the direct sum decomposition of $\cF(\La)$) is
equal to the multiplicity of the weight $\La$ in the irreducible
finite dimensional representation of $\bfg$ with highest weight
$\La+\be$. \par

For $sl(3)$ we have the following
generating function for these multiplicities
\eqn\eqBe{ \eqalign{
\sum_{\be\in Q_+} m(\La;\be) e^\be = &\,
{1\over (1-e^{\al_1})(1-e^{\al_2})(1-e^{\al_3})} \cr
& - {e^{(\La+\rh,\al_1)\al_2}\over (1-e^{\al_2})(1-e^{\al_3})
(1-e^{\al_1+2\al_2})} - {e^{(\La+\rh,\al_2)\al_1} \over
(1-e^{\al_1})(1-e^{\al_3})(1-e^{2\al_1+\al_2})}\,. \cr}
}

Complete reducibility of the $c=\ell$ Fock spaces $\cF(\La)$ follows, as
usual, from the existence of a positive definite hermitian form on
$\cF(\La)$. The rest of the theorem is proved by the standard construction
of a set of singular vectors in $\cF(\La)$ using
screening operators. Completeness of this set then follows
by comparing the characters on both sides of \eqBd.

In the course of the cohomology computation we will need to know
resolutions of the  irreducible $\cW$-modules $L(\La)$.
Completely degenerate modules occur for $\al_+^2 = p'/p$, where
$p,p'$ are two relatively prime positive integers. If we label $\La$
through $\La=\al_+\Lap + \al_-\Lam$, then the set of completely
degenerate modules given by $\Lap\in P_+^{p-\dCo},\, \Lam\in P_+^{p'-\dCo}$
constitute the spectrum of
the so-called $\cW$ minimal models. [Here, $P_+^k$ denotes the
set of integrable weights of level $k$ of the affine Lie algebra $\whg$.]

Resolutions of
the $c<\ell$ minimal modules $L(\Lap,\Lam)$ in terms of Fock spaces were
conjectured in [\BMPb,\FKW]. They can be obtained by performing a
quantum Drinfeld-Sokolov reduction on the corresponding free field
resolution for the underlying affine Lie algebra $\whg$. We will not
repeat them here. For the resolutions
in terms of Verma modules we propose
\thm\thBd
\proclaim Conjecture \thBd. Let $L(\Lap,\Lam),\,\Lap\in P_+^{p-\dCo},
\Lam\in P_+^{p'-\dCo}$ be an irreducible $\cW$ minimal model module. We
have a resolution $(C^{(i)}L(\Lap,\Lam), d'),\,i\leq0$, of
$L(\Lap,\Lam)$ with terms
$$
C^{(i)}L(\Lap,\Lam)\ \cong \ \bigoplus_{ \{w\in\wW\,|\,
\ell(w) = -i\} } \ M(\al_+(w(\Lap+\rh)-\rh) + \al_-\Lam)\,.
$$
\par

Note that the validity of Conjecture \thBd\ is not at all obvious. First
of all, there will be singular vectors
in $M(\La)$ beyond the ones that are used to build the resolution of
Conjecture \thBd. Secondly, since the generator $W_0$ will in general not
be diagonalizable on Verma modules [\Wa],
there is no {\it a priori} reason why all
the terms in the resolution should consist of modules that are induced
from one-dimensional representations of the abelian subalgebra $\{L_0^{(i)}\}$,
\ie Verma modules. Nevertheless, the various examples we have gone through
numerically were not in contradiction with the above conjecture.

For $c=\ell$ irreducible modules $L(\La)$ the situation is rather more
problematic.
Naively, one might still expect Verma module
resolutions analogous to the one in Conjecture \thBd\ (apart from the fact
that now the resolution runs over the finite Weyl group).
This naive guess turns out to be wrong, precisely because of the
problems mentioned above.   A study of examples suggests that one can
still construct resolutions of $L(\La)$, but now including
generalized Verma modules (modules induced from nontrivial representations
of the subalgebra $\{L_0^{(i)}\}$).
We will return to this in Section 4 and [\BMPd].

  In contrast, the $c=\ell$ analogue of the Fock space resolution seems
straightforward
\thm\thBe
\proclaim Conjecture \thBe.
Let $\La\in P_+$. We have a resolution $(C^{(i)}L(\La),
d')$ of the $c=\ell$ irreducible module $L(\La)$ with terms
$$
C^{(i)}L(\La)\ \cong\ \bigoplus_{ \{w\in W\,|\, \ell(w)=-i\} }
\ \cF(w(\La+\rh) - \rh)\,.
$$
\par

\newsec{$\cW_3$ gravity coupled to $\cW_3$ minimal models}

{}From now on we will restrict the discussion to the  $\cW_3$ algebra
generated by $L_n$ and $W_n=L_n^{(3)}$, $n\in\ZZ$.%
\foot{Since, for the most part, we will be using only generic properties
of the $\cW_3$ BRST operator we expect most of the results to generalize
immediately to other $\cW$-algebras.}
Let $V^M$ and $V^L$ be two arbitrary positive energy $\cW_3$ modules,
and let $\cF^{gh(1)}$ and $\cF^{gh(2)}$ denote the Fock
space of a set of first order anticommuting (ghost) fields $(\bgh{1},\cgh{1})$
and $(\bgh{2},\cgh{2})$ of conformal dimensions $(2,-1)$ and $(3,-2)$,
respectively.

Consider the BRST operator $d=  \oint {dz\over2\pi i}J(z)$
acting on $V^M\otimes V^L\otimes
\cF^{gh(1)}\otimes \cF^{gh(2)}$, where [\BLNWa,\BSSBG]
\eqn\eqCc{ \eqalign{
J = & \,\cgh{2}( {\textstyle{{1\over\sqrt{\be^M}}}} W^M
- {\textstyle{ {i\over\sqrt{\be^L}} }}W^L) +
\cgh{1} (T^M + T^L + \half T^{gh(1)} + T^{gh(2)} ) \cr
& + (T^M - T^L) \bgh{1}\cgh{2}\p \cgh{2} +
\mu \p \bgh{1} \cgh{2}\p^2 \cgh{2} + \nu \bgh{1}\cgh{2} \p^3
\cgh{2} \,.\cr}
}
Here $\mu = {3\over5} \nu = {1\over 10 \be^M}(1-17\be^M)$, while
$\be = 16/(22+5c)$.
The operator is nilpotent provided $c^M+c^L=100$, so that on defining
$\al_\pm=\half (\al_0^M\mp i\al_0^L)$ we have $\al_+\al_-=-1$.
The cohomology will be
denoted by $H(d,V^M\otimes V^L)$. It   is graded by
ghost number $\gh{\cdot}$, where $\gh{\cgh{1}}=\gh{\cgh{2}}=-\gh{\bgh{1}}=
-\gh{\bgh{2}}=1$, and the ghost number of the physical vacuum $|0\rangle_{gh}$
(annihilated by all positively-moded ghost oscillators, as well as by the
zero modes $b_0^{[1]}$ and $b_0^{[2]}$)
in the ghost sector is chosen to be zero.

In the sequel,
the module $V^L$, representing the $\cW_3$ gravity (`Liouville')
sector, will be assumed
to be a Fock module $\cF(\La^L,\al_0^L)$, while for the matter sector
we will be interested in either irreducible modules $L(\La)$ or Fock
modules $\cF(\La^M,\al_0^M)$. Our strategy will be to reduce the
computation to that of the cohomology on a product of a Verma
and contragradient Verma module. In the $\cW_3$ gravity sector we will
therefore restrict to states satisfying
$-i(\La^L+\al_0^L
\rh)\in\et^L D_+$, and apply Corollary \thBb\ (b). In the matter sector
we will represent irreducible modules $L(\La)$ by their resolution in terms
of Verma modules (Conjecture \thBd), and compute the BRST cohomology
using the resulting double complex (see,
\eg [\BMPc] for some background material).  In the case of $c^M=2$ Fock
modules, we will first use the decomposition of the Fock space
into irreducible modules (Theorem \thBc), and then proceed as above.

A few comments are in order.
To compute analogous cohomologies $H(d,V^M\otimes V^L)$
in the Virasoro or affine Lie algebra case it often turns out to be convenient
to use the
triangular decomposition $\bfg \cong \bfnp\oplus\bfh\oplus\bfnm$, and pass
to the cohomology relative to the Cartan subalgebra $\bfh$. This relative
cohomology $H(\bfg,\bfh;V^M\otimes V^L)$ can then be studied by means
of a spectral sequence whose first term is given by
(see \eg [\BMPc] and references therein for more details)
\eqn\eqCa{
E_1\ \cong\
\left( H(\bfnm,V^M) \otimes H(\bfnp,V^L)\right)_{\bfh} \,.
}
The absolute cohomology is then eventually
recovered from the relative one by a long exact sequence.

This procedure, however, does not work for $\cW$-algebras for a variety
of reasons. First of all, due to the nonlinear terms in their defining
relations, $\cW$-algebras do not have, strictly speaking,
a triangular decomposition.
As a consequence there does not exist a BRST operator
corresponding to some nilpotent subalgebra, and hence the `splitting'
\eqCa\ cannot make sense for $\cW$-algebras. Also, the passage to the
cohomology relative to the abelian subalgebra generated by $\{ L_0^{tot},
W_0^{tot}\}$, where $L_0^{tot} = \{d,\bgh{1}_0\},\,
W_0^{tot} = \{ d,\bgh{2}_0\}$, is problematic. Although the subset
of states annihilated by $L_0^{tot}, W_0^{tot}, \bgh{1}_0$ and
$\bgh{2}_0$ do form a subcomplex, the complicated expression for
$W_0^{tot}$  clearly makes it hard to determine this subcomplex
explicitly (\ie to write down a basis of states), and it is therefore
unsuitable for calculations. And, even if one could determine this
relative cohomology, it is not clear how to deduce from it the
absolute cohomology due to the nondiagonalizability of $W_0^{tot}$. It is,
for instance, no longer obvious that the absolute cohomology should be
concentrated on the states annihilated by $W_0^{tot}$.%
\foot{They have to be `generalized zero eigenstates' of $W_0^{tot}$,
though, \ie there  should exist an $N$ such that
$(W_0^{tot})^N |\ps\rangle = 0$.}
However, one can still consider the cohomology relative to $L_0^{tot}$ --
a step which in fact is necessary as it effectively reduces the computation
to one on a finite-dimensional complex.
Note that the problems above are somewhat reminiscent of those
occurring for the semi-infinite cohomology of the super-Virasoro
algebra.

As outlined before, the computation of the cohomology
$H(d,L(\Lap,\Lam)\otimes \cF(\La^L,\al_0^L))$ in the case of
states satisfying $-i(\La^L + \al_0^L\rh) \in\et^L D_+$
can be reduced to a computation
of the cohomology on (contragradient) Verma modules, for which we
have the following technical result
\thm\thCa
\proclaim Theorem \thCa. The cohomology $H(d,M(\La^M)\otimes M(\La^L)^*)$
is nonvanishing iff $w(\La^M+\al_0^M\rh)=-i(\La^L+\al_0^L\rh)$ for some
$w\in W$, in which case it is spanned by $\{ v, \cgh{1}_0v,
\cgh{2}_0v, \cgh{1}_0\cgh{2}_0v\}$
where $v$ denotes the highest weight vector of $M(\La^M)\otimes M(\La^L)^*$.
\par

Despite the apparent difficulties in dealing with $\cW$ algebras
mentioned above, it
turns out that the proof of Theorem \thCa\ is not too difficult, and
can be given in complete analogy to the standard proof of \eg
$H(\bfnm,M)\cong \CC$ in the Virasoro or affine Lie algebra case. Namely, by
introducing a filtration whose corresponding spectral sequence
reduces to the Koszul complex in the first term and collapses after that
({\it cf.} \eg [\Knapp]).

One can easily verify that
the `quartet' structure of Theorem \thCa\ will persist in the
cohomology $H(d,L(\Lap,\Lam)\otimes M(\La^L)^*)$ by examining the
zig-zag procedure in the double complex of BRST cohomology and resolution
of $L(\Lap,\Lam)$. In the rest of the paper we will therefore
adopt the terminology `prime state'
for the lowest ghost number state of each quartet
of states of ghost numbers $(G,G+1,G+1,G+2)$, and only mention
results for these prime states, leaving it implicitly understood that
to every prime state there is an associated quartet.\foot{We have adopted
the terminology `prime state' from [\Popetalb], where it was introduced in
a different context.}

Now, restricting the Liouville momentum to $-i(\La^L+\al_0^L\rh)\in
\eta^L D_+$, using Corollary \thBb\ and using the resolution of
 Conjecture \thBd\
for $L(\Lap,\Lam)$, Theorem \thCa\ gives the following result
\thm\thCb
\proclaim Theorem \thCb. For $\al_\pm = \half(\al_0^M \mp i\al_0^L)$, consider
$-i(\La^L+\al_0^L\rh)\in\et^L D_+$, where $\et^L= {\rm sign}(-i\al_0^L)$. Then
$H(d,L(\Lap,\Lam)\otimes \cF(\La^L,\al_0^L)) \neq 0$ iff there
exists a $w\in\wW$ and a $w'\in W$ such that
$w'( (\al_+ w(\Lap+\rh) + \al_-(\Lam+\rh)) =- i(\La^L+\al_0^L\rh)$, in
which case there is a (unique) prime state of ghost number $-\ell(w)$.

The number of affine Weyl group elements with a specified length
$\ell$, or alternatively the number of prime physical states with
ghost number $-\ell$, can be read off from the Poincar\'e series
of $\widehat{W}$ (see \eg [\Hu]) which for $\widehat{sl(3)}$ reads
\eqn\eqCb{
P(t)\equiv \sum_{w\in\wW} t^{\ell(w)} = {(1+t+t^2)\over (1-t)^2} =
1 + 3\sum_{n\geq1} nt^n\,.
}
The level at which this prime state occurs is given by
$(\Lap+\rh - w(\Lap+\rh), \Lam+\rh)$.

In particular, if we take the identity representation $1 \cong L(0,0)$ at
$c^M=0$ (\ie $p'=4,\, p=3$), the BRST operator \eqCc\  reduces to the one
of [\TM], which has been used in recent discussions of two-scalar
gravity (pure $\cW_3$ gravity) [\Popetalb]. In this case Theorem
\thCb\ yields
one $G=0$ prime state at energy level 0,
three $G=-1$ prime states at energy levels $1,1,2$, six $G=-2$ prime states
at energy levels $3,3,4,4,5,5$, nine $G=-3$ prime states at energy
levels $4,6,6,8,8,9,9,11,11$, and so forth. Some of these states have been
explicitly constructed in [\Popetalb]. As a check on Conjecture
\thBd\ we have explicitly constructed some of the remaining ones.

\newsec{The $\cW_3$ string}

In this section we summarize some results on the cohomology of a tensor
product of two Fock space modules, $\cF(\La^M,\al_0^M)
\otimes \cF(\La^L,\al_0^L)$.
A striking difference between the present problem and its analogue
for the Virasoro algebra is that the direct approach
developed in [\BMPa], which employs a spectral sequence arising from
a natural degree on the set of oscillators in the Fock spaces,
seems to be difficult to implement: although it
is still rather straightforward to identify  a spectral sequence
whose  first term gives the usual Virasoro
semi-infinite cohomology, the subsequent terms of this sequence
are of such complexity that a direct analysis seems impossible without
additional insights  into the structure of the BRST complex.
For that reason we will instead follow the path of Section 3 and
reduce the computation to that of the cohomology of Verma modules.

For generic values of the momenta $\La^M$ and $\La^L$ one expects
to find only level $0$ (`tachyonic')
states in the cohomology [\BLNWa], which
is the analogue of the similar result  for the Virasoro
algebra [\LZs,\BMPa].
More precisely,
we parametrize the momenta $\La^M$ and $\La^L$  by
\eqn\DDbb{\eqalign{
\La^M+\al_0^M\rh &= \al_+\La^{(+)} + \al_-\La^{(-)}\,,\cr
-i(\La^L+\al_0^L\rh) &= \al_+\La^{(+)} - \al_-\La^{(-)}\,.\cr}}
We call them  generic iff there is no positive root
$\al\in \De_+$ such that
\eqn\DDbc{(\La^{(+)},\al)\in\ZZ\,,\quad (\La^{(-)},\al)\in\ZZ\,,\qquad
(\La^{(+)},\al) (\La^{(-)},\al)>0\,.}
Then we have
\thm\thDDa
\proclaim Theorem \thDDa. For generic momenta   $\La^M$ and $\La^L$
as defined above,
$H(d,\cF(\La^M,\al_0^M)\otimes \cF(\La^L,\al_0^L))
\not=0$ iff there exists $w\in W$ such that $w(\La^M+\al_0^M\rh)=
-i (\La^L+\al_0^L\rh)$, in which case it is spanned by the states
$v$, $c_0^{[1]}v$,
$c_0^{[2]}v$ and
$c_0^{[1]}c_0^{[2]}v$,
where $v=|\La^M\,\rangle\otimes |\La^L\,\rangle
\otimes |0\,\rangle_{gh}$ is the physical vacuum.

Let us now restrict to the case of $d=(2,2)$ $\cW_3$ string,
namely set $c^M=2$ (and $\al_\pm=\pm1$).
The structure of the cohomology then simplifies
considerably because of the following observation
\thm\thDa
\proclaim Theorem \thDa. The cohomology spaces $H^{(n)}(d,\cF(\La^M,0)
\otimes \cF(\La^L,\al_0^L))$ carry a fully reducible representation of
$sl(3)$. The $sl(3)$ generators are explicitly given by the zero modes
of a level-1
Frenkel-Kac-Segal vertex operator construction in terms of matter
fields only.

Moreover, if we further restrict the Liouville momenta to the
region $-i(\La^L+\al_0^L\rh) \in \et^L D_+$,
we can combine Corollary \thBb\ and Theorem
\thBc\ and reduce the computation of
$H(d,\cF(\La^M,0)\otimes\cF(\La^L,\al_0^L))$ to that of
$H(d,L(\La)\otimes M(\La^L)^*)$ for the $c=2$ irreducible modules
$L(\La)$, which appear in the decomposition of $\cF(\La^M,0)$.  The
latter cohomology can in principle  be determined as in Section 3
({\it cf.}\ Theorem \thCb) provided one knows resolutions of
$L(\La)$'s in terms of Verma modules.

As we have discussed in Section 2, the natural assumption,
that those resolutions are similar to the ones in terms of Fock
spaces given by Conjecture \thBe,  turns out to be incorrect!
In fact, by assuming
the `naive' resolutions one
would not reproduce all the physical states that have been constructed
explicitly in [\BLNWb], most notably those which by the descent equations
[\Wi,\BLNWb] give rise to the $sl(3)$ currents above.

To clarify this we studied  explicitly singular vectors in $c =2$
Verma modules $M(\La )$ for low lying dominant integral  weights $\La \in P_+$.
Specifically, we  have constructed all singular vectors and determined
their embedding patterns up to level
 eight in the Verma modules $M(\La)$ for $\La$  given by (in Dynkin labels)
$(0,0)$, $(1,0)$, $(0,1)$, $(2,0)$, $(0,2)$ and $(1,1)$. From those
explicit examples, the details of which will be presented in [\BMPd],
the following
picture emerges.  Singular vectors are {\it not}
eigenstates of $W_0$ in general, rather they fall into
indecomposable representations of the subalgebra $\{L_0,W_0\}$.
(The simplest example of this phenomenon
occurs for level one singular vectors  in the Verma module with the
weight $(0,0)$, and has already been discussed in [\Wa].) Thus
one is led to consider `generalized Verma modules.' We will denote
by $M(\La)_\nu$ the generalized  Verma module with the highest weight subspace
spanned by such an indecomposable $\nu$-dimensional representation, with
the $L_0$
eigenvalue $h(\La)$, and the generalized $W_0$ eigenvalue $w(\La)$, \ie
$w(\La)$ is a $\nu$-times degenerate root of the characteristic equation.

Assuming that the submodules of (generalized) Verma modules are generated by
singular vectors, an embedding
pattern  then  determines a resolution of the  irreducible module
$L(\La )$  in terms of (generalized) Verma modules. An independent
consistency
check is provided by examining the resulting characters.

To illustrate this let us consider
resolutions of the irreducible module $L(1,0)$ as an example. [To simplify
the notation we will denote the weights by their Dynkin labels.] The
Fock space resolution of Conjecture \thBe\ reads
\eqn\eqAAaa{
\matrix{0&
\rightarrow&
F(3,2)&
\rightarrow&
\matrix{F(4,0)\cr\oplus\cr F(1,3)\cr}&
\rightarrow&
\matrix{F(0,2)\cr\oplus\cr F(2,1)\cr}&
\rightarrow&
F(1,0)&
\rightarrow&
0\cr}}
while that in terms of (generalized) Verma modules is
\eqn\eqAAa{
\matrix{0&
\rightarrow&
M(3,2)&
\rightarrow&
\matrix{M(1,3)\cr\oplus\cr M(3,2)_2\cr}&
\rightarrow&
\matrix{M(4,0)  \cr\oplus\cr M(2,1) \cr \oplus\cr  M(1,3)_2\cr} &
\rightarrow&
\matrix{M(0,2)\cr \oplus\cr M(2,1)_2\cr}&
\rightarrow&
M(1,0)&
\rightarrow&
0\cr}}

Finally, an analogue of Theorem \thCa\ for generalized Verma modules
holds [\BMPd],
so that the cohomology can now be computed as in Section 3 by considering a
double complex. In all the examples we have considered, one finds
that although (generalized) Verma module resolutions can extend beyond
the (minus) third term, as in \eqAAaa,  the resulting (prime) physical states
have ghost numbers $G$ between $-3$ and 0. The cohomology states at
$G=-3$ form the ground ring of the theory [\Wi].   The following
conjecture gives a lower bound on the number of elements of
the ground ring.

\thm\thDb
\proclaim Conjecture \thDb.
Given a pair of integral weights $(\La^M,-i\La^L)$,
let $w\in W$ be such that $w\La^M\in P_+$. Then there are
ground ring elements $\ph_{(\La^M,-i\La^L)}$ provided
$w\La^M + \be =-i\La^L$ for some $\be\in Q_+$. The multiplicity is given
by $m(w\La^M;\be)$, and the energy level at which it occurs is given by
$\half|\La^L + 2\rh|^2 - \half |\La^M|^2$.

This conjecture follows from the decomposition of the Fock space module
in Theorem \thBc\ and  the fact that each irreducible module $L(\La)$
appears to have the Verma module $M(r_3(\La+\rh)-\rh)$  in its resolution,
which gives rise to a $G=-3$ state. Also all the ground ring elements
at level less than 8 are obtained in this way. It appears that
at levels 8 and higher there are additional generators in the ground ring,
so a complete characterization of all $G=-3$ still remains an open problem.

Among the states above,
in particular we find one state at energy level 4, namely $\ph_{(0,0)}$,
which is the $SL(2,\RR)$ vaccum. Then there are
 six states at energy level 6
\eqn\eqDa{\eqalign{
x_1 & = \ph_{(\La_1,\La_1)} \,,\quad  x_2 = \ph_{(\La_2-\La_1,\La_1)}\,,
\quad x_3 = \ph_{(-\La_2,\La_1)} \,,\cr
y_1 &= \ph_{(-\La_1,\La_2)} \,,\quad y_2=\ph_{(\La_1-\La_2,\La_2)}
\,,\quad y_3=\ph_{(\La_2,\La_2)},\cr}}
which transform in the ${\bf 3} + \bar{{\bf 3}}$ under $sl(3)$.
(Note that $x_1$ is
precisely the state arising from the  resolution \eqAAa.)
The states  $x_1,x_3,y_1,y_3$ were  explicitly constructed in
[\BLNWb]. The remaining two  can be obtained by
acting with the $sl(3)$ generators of Theorem \thDa.\foot{We have also
checked explicitly that, in the notation of [\BLNWb],   $J_{\al_3}$  maps
$x_i$  onto $ \ga_0^i$, even without the addition of BRST exact
terms.}

Consider now the ring of operators corresponding to the
states  $x_1,\ldots,y_3$. By examining the cohomology at level 8
one concludes that, after a suitable normalization, their product
must satisfy a constraint $x_1y_1+x_2y_2+x_3y_3=0$. Assuming that
there are no  vanishing relations other than those dictated
by the absence of cohomology at certain momenta, we find
the ring generated by those six operators to consist precisely
of elements listed in  Conjecture \thDb.   It carries a
representation of $sl(3)$ under which it decomposes into a direct
sum of irreducible representations, every inequivalent irreducible
finite-dimensional representation occurring exactly once (this is
called a `model space' for $sl(3)$). The appearance of this `model space'
seems to have been anticipated in [\MMMO].
This ring is easily seen to be equivalent to the ring of harmonic
polynomials (traceless tensors with both co- and contravariant indices)
on the Euclidean 6-plane [\Zh].

We have also verified that all higher ghost number states in the regime
$-i (\La^L+\al_0^L\rh) \in \et^L D_+$ explicitly
constructed in [\BLNWb] arise from those generalized resolutions.

Finally, let us make some remarks on the general case.
In the Virasoro case there exists an isomorphism (as a Virasoro module)
\eqn\eqDb{
\cF(\La^M,\al_0^M)\otimes\cF(\La^L,\al_0^L)\ \cong\
\cF(\La^{M\prime},\al_0^{M\prime})\otimes \cF(\La^{L\prime},\al_0^{L\prime})
\,,}
whenever the labels are related by an $SO(2,\CC)$ rotation
[\It,\LZs]. As a consequence, the cohomologies on these spaces are isomorphic.
In particular, the Fock space cohomologies for $c^M<1$ can be calculated
from the $c^M=1$ cohomology $H(d,\cF(\La^M,0)\otimes\cF(\La^L,\al_0^L))$
by $SO(2,\CC)$ rotation.
Given two two-scalar Fock $\cW$-modules, their tensor product is not
a $\cW$ module, so a naive generalization of \eqDb\ does not
exist. One could look for more general vector space isomorphisms
that intertwine with the BRST operator (these would necessarily have
to act on the ghost Fock spaces too). We have not found any.
Nevertheless, the explicit construction of one-parameter sets
of physical states [\BLNWb] suggests
that the cohomology $H(d,\cF(\La^M,\al_0^M)\otimes \cF(\La^L,\al_0^L))$
at different values of $c^M$ might still be isomorphic.\medskip

\noindent{\bf Acknowledgements:}
J.M. would like to thank A. Ceresole, M. Frau, and A. Lerda for
discussions at an early stage of this work.
\medskip

\noindent
{\bf Note added:} While finishing this paper
we received [\Sa] where results on $\cW$-cohomologies through
Drinfeld-Sokolov reduction are announced.


\footatend\immediate\closeout\rfile\writestoppt
\baselineskip=14pt{\bigskip\noindent {\bf  References}}%
\bigskip{\frenchspacing%
\parindent=20pt\escapechar=` \input refs.tmp\vfill\eject}\nonfrenchspacing

\vfil\eject\end